# Exploring the Naturalness of Buggy Code with Recurrent Neural Networks

**Jack Lanchantin and Ji Gao**
jjl5sw@virginia.edu, jg6yd@virginia.edu
Department of Computer Science
University of Virginia

## Abstract

Statistical language models are powerful tools which have been used for many tasks within natural language processing. Recently, they have been used for other sequential data such as source code. (Ray et al., 2015) showed that it is possible train an *n-gram* source code language mode, and use it to predict buggy lines in code by determining "unnatural" lines via entropy with respect to the language model. In this work, we propose using a more advanced language modeling technique, Long Short-term Memory recurrent neural networks, to model source code and classify buggy lines based on entropy. We show that our method slightly outperforms an *n-gram* model in the buggy line classification task using AUC.

## 1. Introduction

Natural language is inherently very well understood by humans. There are certain linguistics and structures associated with natural language which make it fluid and efficient. These repetitive and predictive properties of natural language make it easy to exploit via statistical language models. Although the actual semantics are very much different, source code is also repetitive and predictive. Some of this is constrained by what the compiler expects, and some of it is due to the way that humans construct the code. Regardless of why it is predictable, it has been shown that code is accommodating to the same kinds of language modeling as natural language (Hindle et al., 2012).

The language modeling task is defined as estimating the probability of a sequence of words (or tokens). Formally, given a sequence of tokens $S$, a language model attempts to estimate the probability of $S$ occurring in the language via the following equation:

$$P(S) = P(s_1) \prod_{i=2}^{N} P(s_t | s_1, s_2, ..., s_{t-1}) \quad (1)$$

Where the conditional probabilities $P(s_t | s_1, s_2, ..., s_{t-1})$

model the probability of token $s_t$ occurring given all previous tokens $s_1, s_2, ..., s_{t-1}$.

From a distribution such as a language model, we can measure the entropy (see section 3.4), or amount of uncertainty of a character given all previous characters. Using this metric, we can determine particular sequences which are "unnatural" with respect to the language.

Recently, (Ray et al., 2015) showed that it is possible to predict buggy lines of code based on the entropy of the line with respect to a code language model. In this work, the authors proposed a *cache language model*, which is an extension of an *n-gram* language model to handle local regularities in a piece of code which is being examined for bugs. They combine both a global and cache language model to measure the entropy of a certain line. They provide extensive experimentation to show that entropy of code can successfully be used to determine buggy lines similar to, and in some cases better than previous state-of-the-art bug localization tasks.

The main drawback of their paper is that they use a typical *n-gram* language model, which struggles to handle long term dependencies due to the computation cost of a large *n*. Recent work has shown that recurrent neural network (RNN) models are able to model languages with long term dependencies much better than previous techniques such as *n-grams* (Graves, 2013; Sutskever et al., 2014; Sundermeyer et al.; Karpathy et al., 2015).

In this work, we propose using a recurrent neural network, and more specifically, a Long Short-term Memory (Hochreiter & Schmidhuber, 1997) network in order to create a source code lanuage model. At the cost of a harder optimization problem, LSTMs are able to model the full conditional distribution of a sequence. We then use this language model to compare our results with (Ray et al., 2015) in classifying buggy lines based on entropy.

### 1.1. Motivation for Using LSTMs for Modeling Code

Using LSTMs in modeling source code is arguably more important than in the natural language modeling case. It has been shown that although computationally expensive, modest sized *n-gram* models can model languages and generate text which is comparable to human writing. However, this is due to the fact that natural language is fairly local. Most words (or characters) do not heavily depend on words farther than about 20 back. In the source code case, it is drastically different. For example, a function could be 20 lines (and thus > about 200 tokens) long. The characters at the end of the function are heavily dependent on the characters at the beginning of the function, and possibly even lines before the function. Furthermore, there are multiple dependencies within source code in order to model the sequence (e.g. syntax, indenting, variable names). Although it is hard to properly model all aspects of code in a mode, it has been argued that a deep stack of non-linear layers in between the input and the output unit of a neural network are a way for the model to encode a non-local generalization prior over the input space (Bengio, 2009; Szegedy et al., 2013). In other words, it is assumed that is possible for the language model to model regions of the input space (sequences) that contain no training examples, but are interpolated versions in the vicinity of training samples. This is important because code can be highly variable in terms of the actual tokens, but the underlying mechanisms and structure is consistent.

Recently, it was shown by (Karpathy et al., 2015) that we can easily generate source code, which appears to be written by a human, from a source code language model. This is impressive because in the examples shown, the code is well indented, the braces and brackets are correctly nested, and even commented correctly. This is not something that can be achieved by looking at the previous $n$ characters. They found certain LSTM "cells" which were responsible for code semantics such as indentation, if statements, and comments.

In this sense, we hypothesize that LSTMs are better suited to model source code, and thus will give higher accuracy bug predictions.

## 2. Related Work

### 2.1. Bug Detection in Software Engineering

For software bug detection, there are two main areas of research: bug prediction and bug localization.

(1) Bug prediction, or statistical defect prediction, which is concerned with being able to predict whether or not there is a bug in a certain piece of code, has been widely studied in recent years (Catal & Diri, 2009). With the vast amount of archived repositories in websites containing bug reports, there are many opportunities for finding bugs in code.

(2) Bug localization is concerned with exactly locating or classifying specific lines as buggy or non-buggy. Static bug finders (SBFs) automatically find where in code a bug is located. SBFs use properties of code to indicate locations of bugs. There has been a wide array of recent work in this area (Rahman et al., 2014), which use many pattern recognition techniques to find bugs. As noted in (Ray et al., 2015), using SBFs and using the entropy of language models are two very different approaches to achieve the same goal. The main goal of their work is to compare the effectiveness of language models vs SBFs for the same task of classifying buggy lines.

### 2.2. Natural Language Processing

There have been many works in NLP which use language models for sequential tasks such as word completion (Mikolov et al., 2013), machine translation (Bahdanau et al., 2014), and many others. There are also a variety of models which use word embeddings (Mikolov et al., 2013) which are learned from language models in order to perform other tasks such as sentiment analysis.

Separately, there are a variety of models which do not use language models, but do sequence classifications (e.g. sentiment analysis, document classification) using pattern recognition techniques (Zhang et al., 2015; Yang et al., 2016). These would be similar to SBF techniques in software engineering. However, to the best of our knowledge, there have not been any works in NLP which use entropy from neural language models to classify sequences.

## 3. Methods

### 3.1. Recurrent Neural Network Language Models

Recurrent neural networks (RNNs) are models which are particularly well suited for modeling sequential data. At each time step $t$, an RNN takes an input vector $\mathbf{x_t} \in \mathbb{R}^n$ and a hidden state vector $\mathbf{h_{t-1}} \in \mathbb{R}^m$ and produces the next hidden state $\mathbf{h_t}$ by applying the following recursive operation:

$$\mathbf{h_t} = f(\mathbf{Wx_t} + \mathbf{Uh_{t-1}} + \mathbf{b}) \tag{2}$$

The output prediction $\mathbf{y_t} \in \mathbb{R}^n$ is made via the following operation:

$$\mathbf{y_t} = \mathbf{Ch_t} \tag{3}$$

Where $\mathbf{W} \in \mathbb{R}^{m \times n}, \mathbf{U} \in \mathbb{R}^{m \times m}, \mathbf{C} \in \mathbb{R}^{m \times n}, \mathbf{b} \in \mathbb{R}^m$ are the learnable parameters of the model, and $f$ is an element-wise nonlinearity. The parameters of the model are learnt via backpropagation through time (Werbos, 1990). Due to their recursive nature, RNNs can model the full conditional distribution of any sequential distribution. However, RNNs suffer from what is referred to as the "vanishing gradient" problem, where the gradients of time steps far away either vanish toward 0, or explode toward infinity, thus making the optimization of the parameters difficult.

To handle the vanishing gradients problem, (Hochreiter & Schmidhuber, 1997) proposed Long Short-term Memory (LSTM), which can handle long term dependencies by using "gating" functions which can control when information is written to, read from, and forgotten. Specifically, LSTM "modules" take inputs $\mathbf{x_t}, \mathbf{h_{t-1}}$, and $\mathbf{c_{t-1}}$, and produce $\mathbf{h_t}$,

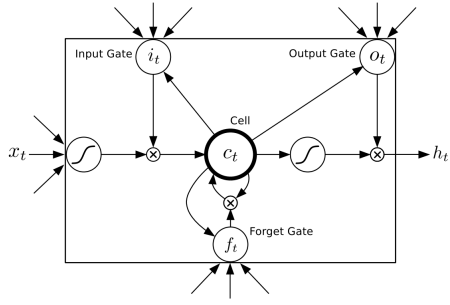

*Figure 1.* An LSTM Module

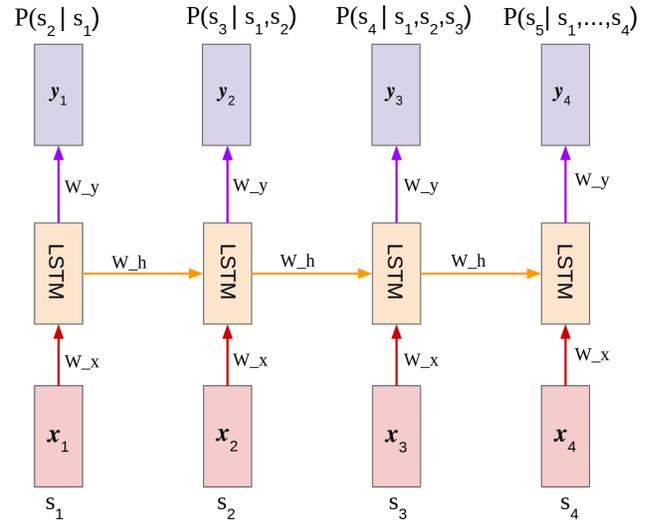

*Figure 2.* Model Overview

and $\mathbf{c_t}$ in the following way:

$$\mathbf{i_t} = \sigma(\mathbf{W^i x_t} + \mathbf{U^i h_{t-1}} + \mathbf{b^i})$$
$$\mathbf{f_t} = \sigma(\mathbf{W^f x_t} + \mathbf{U^f h_{t-1}} + \mathbf{b^f})$$
$$\mathbf{o_t} = \sigma(\mathbf{W^o x_t} + \mathbf{U^o h_{t-1}} + \mathbf{b^o})$$
$$\mathbf{g_t} = tanh(\mathbf{W^g x_t} + \mathbf{U^g h_{t-1}} + \mathbf{b^g}) \qquad (4)$$
$$\mathbf{c_t} = f_t \odot c_{t-1} + i_t \odot g_t$$
$$\mathbf{h_t} = o_t \odot tanh(c_t)$$

Where $\sigma()$ and $tanh()$ are element-wise sigmoid and hyperbolic tangent functions. $\odot$ represents an element-wise multiplication. $\mathbf{i_t}$, $\mathbf{f_t}$, and $\mathbf{o_t}$ are referred to as the input, forget, and output gates, respectively. An overview of an LSTM module can be seen in Figure 1. It is the gating mechanisms that allow the LSTM to remember long term dependencies, which are especially important in code where a certain line may depend on code many lines back.

## 3.2. Handling the Global and Local Representations of Code

One significant difference between natural language and source code is that source code has specialized local dependencies such as variable names. It is hard to capture elements such as variable names with global source code language models. To handle this, we propose a model similar to (Ray et al., 2015), where we construct both a "global" and "local" (i.e. cache) language model, each being an LSTM. Details of how the global and local models are trained is explained in section 4.

## 3.3. Model Details

An overview of our model can be seen in Figure 2. The LSTM modules are as described in section 3.1. In this figure, we show the "unrolled" version of the model, which shows what happens at each time step for 4 different input tokens. Each LSTM module has 128 cells, or units. Although it is shown with only one in the figure, our implementation uses two layers of LSTM modules. That is, there is a second LSTM on top of the LSTM shown in the figure. This stack of LSTMs allows for a better representation of the input to output signal.

$W_x$, $W_h$, and $W_y$, are the learned input, hidden, and output

embeddings of the model, where $W_h$ encapsulates all learned parameters of the LSTM as explained in 3.1. There are multiple ways to represent the input embedding $W_x$. It could be done using a "one hot" embedding, but we use learned character level embeddings (Mikolov et al., 2013) which can learn representations among characters. Our character embeddings are vectors of length 64. Although we do not do so in this work, these embeddings could be used to do other tasks such as classification directly from input sequences.

We use the Adam method of optimization (Kingma & Ba, 2014), using mini-batches of 128 sequences at a time. During training, we optimize over only 50 characters at a time, but it has been shown that LSTMs can still learn dependencies longer than the training sequence length (Karpathy et al., 2015).

We implement our model using the Torch7 framework (Collobert et al., 2011). Our code is available at https://github.com/jacklanchantin/sclm.

## 3.4. Entropy

For any probability distribution, the entropy is defined as the amount of uncertainty in the distribution. Formally, the entropy $H$ is:

$$H = -\sum_{x \in X} p(x)\mathrm{log}p(x), \qquad (5)$$

where the summation is over all values of the random variable $X$. In our case, we are interested in the amount of uncertainty of predicting the next character given all previous characters based on our source code language model. In other words, the distribution is over all possible characters given the previous characters, where $X$ is the dictionary of characters. If the output probability of the next character is close to uniform, then the entropy will be high, meaning that the predicted character is "unnatural" with regard to the language model. In order to determine the entropy of a line of code, we take the average

| | Snapshots | Java files | Lines |
|---|---|---|---|
| Global Training | 50 | 118164 | 16502732 |
| Local Training | 18 | 59180 | 9437902 |
| Total | 68 | 177344 | 25940634 |

*Table 1.* Statistics of the datasets. We use the first 50 snapshots as a global training set, and the other 18 snapshots as a local testing set. We test on 10,902 buggy and 10,902 non-buggy lines from the local training set (which are excluded in the actual training).

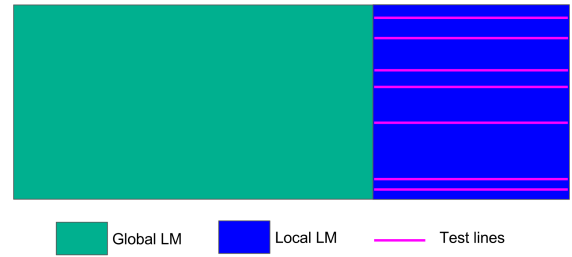

*Figure 3.* Experimental configuration: We use the first 50 snapshots (green) to train a global model. The remaining 18 snapshots (blue) excluding the testing lines are used as the local training set. We select 10,492 buggy and 10,492 non-buggy lines from the training set (pink lines).

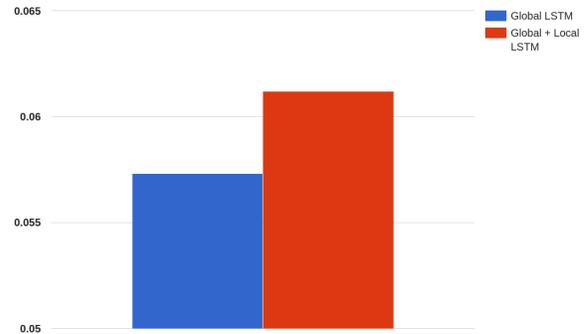

*Figure 4.* Average entropy difference (average entropy of buggy code - average entropy of non-buggy code). The blue bar is the result from Global model, and the red bar is the result from Global+Local model.

entropy of each character in the line.

To handle the global and local language models, we formulate the total entropy as:

$$H_{total} = \lambda H_{global} + \lambda H_{local} \qquad (6)$$

Where $\lambda$ is a weighting term. We use $\lambda = 0.5$ in our experiments. The main contribution of this work is to augment (Ray et al., 2015), which showed that it is possible to use this metric of uncertainty in a line to predict a bug. We simply implement a more advanced language modeling technique, thus the only difference is the distribution $p$.

# 4. Experiments and Results

## 4.1. Dataset

To train and test an LSTM language model which could be used to distinguish buggy and non-buggy code, we choose the ElasticSearch Project on Github, which has revision histories containing comments and commits containing bug line indications.

The ElasticSearch project is a distributed search engine built for the cloud, written in Java. We use 68 snapshots of the project, which in total contain 177,344 Java files and 25,940,634 lines of code. Detailed information of the data set is listed in Table 1.

## 4.2. Experimental Design

To show our model can detect buggy code in a real environment, we separate the dataset into two parts: a global training set and a local testing set. By separating the training data and testing data, we aim to show that our model could be used to predict buggy codes in the development phase of a project.

To construct our experimental setting, the first 50 snapshots are used to form the global training set. The local training set contains other 18 snapshots from the project, excluding the testing lines. To form the test set, we choose all 10,902 buggy lines and an matching number (10,902) of non-buggy lines from the total 9,437,902 lines in the local training set. This ensures that the tested lines have the local semantics of the local language model, but also the global semantics of the global model. An overview of the data is shown in Figure 3.

## 4.3. Model design and Metrics

To model the code, we train a separate LSTM on the Java files in the global and local training sets. From these trained models, we seek to show their modeling capacity by predicting bugs based on the entropy of lines in the testing set with respect to the conditional distribution of the language model.

Our baseline model is the *n-gram* language model from (Ray et al., 2015), which is also used to predict whether a line of code is buggy or not. To compare our model with theirs, we use the average entropy of each line (see section 3.4), and evaluate the positive and negative buggy lines with entropy using area under the ROC curve (AUC). AUC is a traditional non-parametric metric to evaluate the performance of a binary classification result. A larger AUC value suggests that the classification result is well-ordered and less mistaken. AUC is a valid metric because the actual prediction values (i.e. entropy), don't matter, but rather the ordering of entropy taking into account the labels.

*Table 2.* Average entropy value of the 10,492 buggy lines and 10,492 non-buggy lines from our 2 models.

| Model | Buggy Lines | Non-Buggy Lines |
|-------|-------------|-----------------|
| Global LSTM | 1.6498 | 1.5925 |
| Global + Local LSTM | 1.5842 | 1.523 |

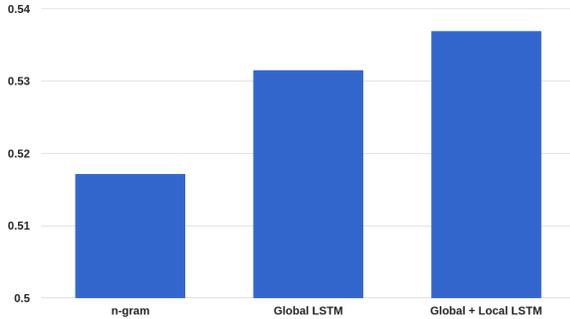

*Figure 5.* Comparing the AUC value of three tested models. Both our models (left 2) achieve a higher AUC than the baseline.

### 4.4. Results

#### 4.4.1. Entropy

Since entropy is relative, we can't directly compare our entropy result with the entropy result from the baseline model. But by showing the difference between the entropy of buggy lines and non-buggy lines, we can justify the validity of our model. This is shown in Table 2 and Fig 4.

From Table 2 we can see that buggy lines generally have a larger entropy value, which accords our assumption. Also, from Figure 4, we can see that the mixed global + local model has a larger average entropy difference than just the global. This proves that the local model can help in capturing some of the local regularities in the project.

#### 4.4.2. AUC

Figure 5 shows a comparison of AUC values for the three models. We can see from the graph that both our global LSTM and global + local LSTM achieve a slightly larger AUC value than the baseline language model, which illustrates that our model does a better job capturing the true semantics of the non-buggy lines.

## 5. Threats to Validity

The main threat to the validity of our experiments are the datasets used. We believe that the training set and testing set are too similar, leading to an over-fitted language model. Similarly, the local vs global datasets are too similar, not allowing for a better opportunity to capture local regularities. Also, there are some buggy lines in the training set, but we believe that since there is only a very small percentage of them, they are essentially disregarded in the language model. See section 6 for our proposal of future work which we believe could alleviate the issues proposed.

The second threat is that we use Area Under ROC Curve (AUC) scores to evaluate our predictions, where it has been shown that the Area Under the Cost-Effectiveness Curve (AUCEC) is the optimal metric for evaluating bug localization (Arisholm et al., 2010). For future work, we would like to evaluate our method using AUCEC.

It is difficult to directly compare our results vs the results in (Ray et al., 2015) due to the fact that entropy is relative, so the values do not matter. We attempted to handle this issue using AUC.

Lastly, it is hard to say that this technique is generalizable since we only used one small dataset. However, the main goal of this work was to show that more advanced language modeling techniques are applicable for such software engineering tasks. We hope to further validate our approach by applying our method on a greater number of datasets.

## 6. Conclusion and Future Work

In this work, we show that a more advanced language model, namely a Long Short-term Memory (LSTM) recurrent neural network, can outperform a simpler $n$-$gram$ model for the task of buggy line classficiation. We hypothesize that LSTMs are able to capture better regularities in the source code than previous techniques.

For future work, we would like to expand on this idea and implement our techniques on a new dataset. We believe that the best way to evaluate this model and technique is to implement the following two steps: (1) train a language model on many different projects of the same language (e.g. Java). This will create a better "true" global language model which captures the regularities of the language semantics itself. (2) Train a separate language model on just one project that you are interested in predicting bugs, regardless of snapshot time. This captures a better version of the local semantics. We would also like to explore different parameters of the global vs local weighting term $\lambda$.